# Linking Data Citation to Repository Visibility: An Empirical Study

Fakhri Momeni[1], Janete Saldanha Bach[1], Brigitte Mathiak[1] and Peter Mutschke[1]

[1] *firstname.lastname@gesis.org*

Knowledge Technologies for the Social Sciences (KTS), GESIS - Leibniz Institute, Unter Sachsenhausen 6-8
50667 Cologne, Germany

**Abstract**

In today's data-driven research landscape, dataset visibility and accessibility play a crucial role in advancing scientific knowledge. At the same time, data citation is essential for maintaining academic integrity, acknowledging contributions, validating research outcomes, and fostering scientific reproducibility. As a critical link, it connects scholarly publications with the datasets that drive scientific progress. This study investigates whether repository visibility influences data citation rates. We hypothesize that repositories with higher visibility, as measured by search engine metrics, are associated with increased dataset citations. Using OpenAlex data and repository impact indicators (including the visibility index from Sistrix, the h-index of repositories, and citation metrics such as mean and median citations), we analyze datasets in Social Sciences and Economics to explore their relationship. Our findings suggest that datasets hosted on more visible web domains tend to receive more citations, with a positive correlation observed between web domain visibility and dataset citation counts, particularly for datasets with at least one citation. However, when analyzing domain-level citation metrics, such as the h-index, mean, and median citations, the correlations are inconsistent and weaker. While higher visibility domains tend to host datasets with greater citation impact, the distribution of citations across datasets varies significantly. These results suggest that while visibility plays a role in increasing citation counts, it is not the sole factor influencing dataset citation impact. Other elements, such as dataset quality, research trends, and disciplinary norms, can also contribute to citation patterns.

**Introduction**

In the ever-evolving landscape of scholarly research, the effective citation of data is fundamental in fostering transparency, reproducibility, and the robust advancement of scientific knowledge. As data-intensive research continues to expand, the growing volume and diversity of available datasets make standardized and effective data citation practices increasingly crucial.

This study examines the relationship between repository visibility, repository impact, and dataset citation rates. We define repository visibility through Sistrix's visibility index, which measures how discoverable a repository is via search engines, and repository impact through the h-index of datasets hosted on a domain. The "Joint Declaration of Data Citation Principles" by the Data Citation Synthesis Group (Martone, 2014) represents a key contribution from the FORCE11 community that organized the FAIR data principles. By emphasizing standardized, accessible, and transparent data citation practices, the declaration provides a foundation for this research. Building upon such works, this study aims to offer detailed insights into the interplay between repository visibility and data citation, addressing critical needs in the evolving scholarly landscape.

Lin et al. (2014) initiative underscores the ongoing efforts to establish metrics that quantify data usage and citation, thereby recognizing and attributing credit to data authors for their contributions. Moreover, studies like that of Piwowar and Vision (2013) highlight the potential advantages of open access to data, suggesting a correlation between openly accessible datasets and increased citation rates.

This paper examines the connection between repository visibility (Sistrix index) and repository impact (h-index of hosted datasets) in relation to data citation rates. It also addresses the challenges highlighted by Starr et al. (2015) regarding the accessibility of cited data in scholarly publications. Furthermore, Robinson-Garcia et al. (2017) evaluate DataCite as a bibliometric source, offering insights into the tools available for analyzing data citation trends.

Navigating the general agreement and debates surrounding data publication, as discussed by Kratz & Strasser (2014), this study contributes to the ongoing conversation by quantifying the relationship between repository visibility, repository impact, and dataset citation. By integrating these perspectives, we aim to refine our understanding of how dataset discoverability and repository influence shape citation practices.

We are particularly interested in the following research question: Does the visibility of data repositories, as measured by the Sistrix index, correlate with their citation impact (h-index, mean, and median citations) and the number of citations received by datasets they host?

**Related work**

Several studies have looked at the challenges of citing data to make it easier to find and access. For example, Krause & Mongeon (2023) examined how datasets are cited in the OpenAlex database. They found interesting patterns in how data creators are connected to the authors who cite their work. Their study revealed trends in citation across different countries and emphasized the importance of open research practices for clear and transparent scholarly work.

Park, You & Wolfram (2018) highlighted the prevalence of informal data citation within scientific publications, particularly in the biological and biomedical sciences, the fields with the most public data sets available documented by the Data Citation Index (DCI). Their study underscored the challenges in adequately acknowledging and documenting data contributions alongside formal data citation practices, emphasizing the need for streamlined citation methodologies to encompass informal data attributions effectively.

Addressing the imperative for FAIR principles (findability, accessibility, interoperability, reusability) in biomedical datasets, Tsueng et al. (2023) highlighted the challenges associated with achieving FAIRness. Their findings emphasized the absence of a unified metadata standard among repositories housing these datasets, hindering the discoverability and accessibility of datasets on major platforms such as Google Dataset Search. This underscores the importance of metadata standardization in improving dataset accessibility and visibility.

Innovative approaches proposed by Bach, Klas & Mutschke, (2022) introduced an infrastructure to assign Persistent Identifiers (PIDs) to dataset elements (i.e., variables) within Social Sciences datasets. This novel framework tackled data citation and reuse obstacles by offering a structured method to reference specific elements within data files, thus facilitating retrieval with requisite metadata. This approach catered to both machine-actionable and human-accessible needs, significantly improving data citation practices in the Social Sciences fields.

Groth et al. (2020) emphasized the pivotal role of data citation and its underlying infrastructures, particularly associated metadata, in enabling FAIR data reuse. This paper underscored the importance of data citation in rendering datasets findable and accessible. It advocated for consistently implementing and supporting machine-readable metadata to address challenges hindering maximal and appropriate reuse of existing datasets.

Onyancha (2016) found a strong correlation between data citation and article citation (correlation coefficient of 0.68) as well as between data citation and the h-index of journals (correlation coefficient of 0.71). These findings highlight the interconnectedness between data citation practices and the scholarly impact of articles within their respective journals.

While these studies have significantly contributed to understanding data citation dynamics and enhancing data accessibility, there remains a notable research gap concerning the direct impact of repository visibility and data access status on data citation rates. This study aims to investigate the intricate relationship between data citation, repository findability, and data access in the Social Sciences and Economics fields, aiming to elucidate how these factors influence citation rates. By addressing these factors, the study seeks to benefit researchers by promoting proper data citation practices, which can enhance research visibility, foster academic credibility, and ensure acknowledgment of contributors' work. Additionally, improving data citation practices benefits the broader research community by facilitating transparency, reproducibility, and collaboration. This makes data more discoverable and accessible, directly appealing to researchers as key stakeholders who rely on well-documented and accessible datasets.

The central research question involves uncovering how much repository visibility and data access status correlate with data citation rates, thereby highlighting the critical need to engage researchers and repository managers in adopting practices that improve data sharing and citation.

**Data and Methods**

*Data Source and Dataset Selection*
We utilized the 2023 version of the OpenAlex database, maintained by the German Competence Centre for Bibliometrics, to gather bibliometric information on published datasets. It offers openly available metadata on scholarly works, including publications, datasets, authors, institutions, journals, and research topics, facilitating comprehensive analyses of research outputs and their impact. We identified datasets by filtering those where the *'item_type'* value was *'dataset'* within the table containing published items in the database. From the expansive collection of datasets covering 284 subject categories, our focus centered on datasets associated with the following fields: Economic[1] and Social Sciences[2].

Our dataset compilation comprised 401,659 datasets in the 'Social Sciences' category and 78,267 datasets in the 'Economics' category.

There is an overlap involving 37,160 datasets in the 'Social Sciences' and 'Economics' categories.

A notable observation emerged when analyzing the citation rates of datasets within these fields in OpenAlex. Surprisingly, our findings revealed that most datasets in these fields had no citations recorded in OpenAlex. Specifically, within the 'Social Sciences' category, a staggering 99% of datasets had no recorded citations, while in the 'Economics' category, 98% of datasets remained uncited. This remarkable prevalence of uncited datasets raises critical questions about the relevance of these datasets, the extent to which they are reused in scholarly research, and whether they are being utilized without proper citation, highlighting a significant challenge in promoting data attribution and recognition within academic work.

We extracted digital object identifiers (DOIs) from datasets listed in OpenAlex and determined their primary web domains using a Python script.

*Key Measures: h-Index and Visibility Index*
To explore the factors associated with dataset citation rates, we focus on two key measures: the visibility index and h-index. These metrics were selected for their complementary roles in capturing the scholarly impact and visibility of web domains hosting datasets. By investigating these measures, we aim to uncover patterns that might inform strategies for improving dataset attribution and visibility.

---

[1] It includes Development Economics, Economic Geography, Financial Economics, Law and Economics, Economic History, Economic System, Environmental Economics, Keynesian Economics, Econometrics, Labour Economics, Natural Resource Economics, Socioeconomics, Neoclassical Economics, Demographic Economics, Mathematical Economics, Welfare Economics, Political Economy, Economic Policy, Economic Growth, Market Economy, Development Economics, Public Economics, Macroeconomics, Monetary Economics, International Economics, Agricultural Economics, Economic Geography, Positive Economics, Classical Economics, Economy, Finance, Financial System, and Business Administration

[2] Encompassing disciplines such as Social Science, Archaeology, Anthropology, Developmental Psychology, Law, Library Science, Linguistics, Political Economy, Communication, Demography, Gender Studies, Public Relations, Public Administration, Social Psychology, Socioeconomics, Pedagogy, Management Science, and Management

The visibility index, obtained from Sistrix[3], is widely used in digital marketing and search engine optimization. It measures how visible a web domain is in search engine results, offering insight into its discoverability. It is done in three steps: collection of data, weighting of data and summation of the values for the visibility index. A detailed explanation of this methodology is available in Sistrix's official documentation[4]. In our study, we compute a web domain's visibility index by averaging its scores across 30 countries, as visibility plays a crucial role in dataset findability and potential citation impact. To retrieve visibility index values, we utilized the *Sistrix API*[5] and developed a script that queries the visibility scores for each domain across multiple country-specific search engine indexes. This automated approach ensures consistency and scalability in our data collection process. The full code is available at: GitHub Repository[6]. Traditionally, the h-index has been used to measure the impact of authors or journals based on citation data (Mester, 2016). In this study, we extend its application to measure the impact of web domains hosting datasets. This novel adaptation evaluates how influential a domain is in facilitating impactful research through its datasets.

The h-index of a web domain is calculated by ranking its datasets in descending order based on the number of citations and finding the point where the rank equals or exceeds the number of citations. Mathematically:

$$h = \max\{h' : h' \leq c_{h'}\}$$

where $h'$ is the rank of a dataset, and $c_{h'}$ is the number of citations of the dataset ranked $h'$. For example, a domain with an h-index of 20 has at least 20 datasets, each cited 20 or more times, while all other datasets have fewer than 20 citations.

This adaptation leverages the h-index's ability to account for both the quantity (number of datasets) and the quality (number of citations) of datasets hosted on a web domain, providing a balanced metric of impact. Similar to its application in author-level metrics, the h-index for web domains highlights the extent of reuse and scholarly influence of the datasets they host.

*Data Filtering and Analysis*
To focus on datasets that reflect current trends and practices in data citation and visibility, we excluded datasets published before 2016. This decision ensures that the analyzed data remains relevant and aligns with the dynamic nature of the visibility index and the evolving landscape of research dissemination. Consequently, we applied a filtering process to identify newer datasets published from 2016 onward. This systematic approach led us to analyze a total of 155,564 datasets in Social Sciences and 37,621 in Economics, with an overlap of 18,998 datasets between the two fields.

The number of datasets associated with each web domain, their respective average visibility index (have been acquired from Sistrix in August and September 2023), and the computed h-index for each web domain (based on the citation counts of datasets hosted on the web domain) can be accessed on GitHub.

---

[3] https://www.sistrix.com/api/domain/domain-visibilityindex/
[4] https://www.sistrix.com/visibility-index/calculation
[5] https://www.sistrix.com/api/domain/domain-visibilityindex/
[6] https://github.com/momenifi/Dataset_finability/blob/main/2visibility_domains.py

We computed the mean and standard deviation for the h-index and visibility index to offer statistical insights into the data's distribution and central tendencies. We also computed the mean and standard deviation of normalized citation[7] for the datasets published during 2016 and 2023 (155,564 in Social Sciences and 37,621 in Economics). **Table 1** displays the mean and standard deviation. The notably high standard deviation observed in both the Social Sciences and Economics categories indicates considerable variability within the dataset.

*Correlation Analysis*
Given the high variability in citation counts and other metrics, we employed Spearman's coefficient to determine the correlation between these variables. Spearman correlation, as a non-parametric measure, evaluates the strength and direction of monotonic relationships rather than relying on the actual values, rendering it less sensitive to extreme values or outliers. Given its resilience to extreme values, Spearman's correlation is a more suitable metric for depicting the monotonic relationship between these variables.

**Table 1. Mean and Standard Deviation (SD) for the h-Index and the Visibility Index of web domains, and the number of citations received by datasets.**

| *Variable* | *Mean (Social Sciences)* | *SD (Social Sciences)* | *Mean (Economics)* | *SD (Economics)* |
|---|---|---|---|---|
| *h-index* | 0.95 | 5.1 | 0.89 | 4.84 |
| *Visibility Index* | 0.22 | 3.03 | 0.32 | 3.9 |
| *Normalized citation* | 0.08 | 2.07 | 0.14 | 1.76 |

**Results**
We investigated how web domain visibility indices relate to the citation impact of their respective datasets. To gauge the citation impact within each web domain, we calculated the eight-year h-index for each web domain using published datasets from 2016 to 2023, focusing on the *Social Sciences* and *Economics* categories. The eight-year h-index for a web domain, denoted as $h$, indicates that the web domain $d$ should have at least $h$ datasets published between 2016 and 2023, with each dataset receiving a minimum of $h$ citations.

Tables *A.1, A.2, A.3, A.4, A.5*, and *A.6* present the top ten web domains based on the number of datasets, web domains' h-index, and visibility index. **Table A.5** and **Table A.6** illustrate the ten most influential web domains (in terms of the h-index) under which datasets have been registered. Notably, not all these web domains are traditional data repositories. Brill-online is a collection of annotated historical texts, some of which are highly relevant to Social Sciences and Economics theory, and Psycnet is mainly a journal for Psychology, which also stores other entity types, such as instruments, which are often used in questionnaires used to conduct surveys. While these instruments are categorized as datasets in the metadata, they are not what one typically envisions when talking about datasets, just like the historical texts. However, we would argue they are in the same category as datasets, in that they are resources produced by researchers for other researchers to enable or facilitate research, and they follow the same rules. Additionally, both annotated texts and instruments are both commonly referred to as "data" in their respective home disciplines: literary studies and psychology. We see this as an example of how research data infrastructures designed for one discipline elevate research in other disciplines as well.

---
[7] by dividing the number of citations by the age of publication (age of publication is equal to 2023 - publication year)

There are some systematic biases in play as well. ICPSR is one of the largest data providers for quantitative data. However, there are only a handful of citations registered in the dataset. We know that citation of data is often done informally (Boland et al., 2012), therefore we assume that the use of these datasets is strongly underreported.

The **Table** *2* and **Table** *3* summarize the relationship between the visibility index of web domains and their citation impact in the fields of *Social Sciences* and *Economics*. These tables present the Spearman correlation coefficients for different h-index threshold levels, illustrating how the strength of association between visibility and impact changes across subsets of web domains. In addition to the h-index, the analysis also includes mean and median citations, providing a more detailed view of how web visibility relates to different aspects of citation impact.

The results indicate a positive correlation between visibility and the h-index across all thresholds, with stronger associations observed at higher h-index levels. This suggests that domains with greater web visibility tend to be linked to more impactful research, particularly among the most highly cited domains. However, the correlation between visibility and citation counts (both mean and median) is weaker and inconsistent. In Social Sciences, mean citations show a slight positive correlation at the broadest threshold but turn negative at higher h-index thresholds, while median citations exhibit no clear pattern. In Economics, visibility has little to no correlation with citation counts, suggesting that highly cited research does not necessarily originate from highly visible domains.

These findings reinforce the idea that web visibility is associated with domain-level research impact, as measured by the h-index, but does not directly translate into higher citation counts. While visibility may enhance discoverability, other factors (such as dataset quality, research trends, and disciplinary practices) play a crucial role in shaping citation impact.

Table 2. Spearman Correlation between h-index of web domains, dataset's citation metrics (mean and median citation) in *Social Sciences* and visibility index of web domain captured by Sistrix. The table presents correlation values along with their respective p-values in parentheses.

| h-index Threshold | Number of Web Domains | h-index Correlation | Mean Citations Correlation | Median Citations correlation |
|---|---|---|---|---|
| All web domains | 389 | 0.14 (0.005) |  | -0.043 (0.393) |
| Web domains with h-index > 0 | 144 | 0.23 (0.004) | -0.112 (0.181) | -0.147 (0.078) |
| Web domains with h-index > 1 | 55 | 0.37 (0.005) | -0.002 (0.986) | -0.004 (0.975) |

Table 3. Spearman Correlation between h-index of web domains, dataset's citation metrics (mean and median citation) in *Economics* and visibility index of web domain captured by Sistrix. The table presents correlation values along with their respective p-values in parentheses.

| h-index Threshold | Number of Web Domains | h-index Correlation | Mean Citations Correlation | Median Citations correlation |
|---|---|---|---|---|
| All web domains | 225 | 0.14 (0.040) | 0.097 (0.148) | 0.008 (0.908) |
| Web domains with h-index > 0 | 84 | 0.31 (0.004) | 0.0001 (0.999) | -0.055 (0.622) |
| Web domains with h-index > 1 | 25 | 0.47 (0.017) | -0.028 (0.895) | -0.052 (0.805) |

**Table *4*** and **Table *5*** [8] provide additional insights by examining correlations at the dataset level. The results show that visibility correlates more strongly with whether a dataset receives at least one citation rather than with citation counts beyond the first citation. This aligns with prior research emphasizing that literature plays a key role in helping researchers discover datasets (Gregory et al., 2020). The correlation between visibility and first citation suggests that repository visibility plays an essential role in the initial citation of datasets, but other factors such as dataset quality, disciplinary norms, and research trends may be more influential in determining long-term citation impact.

Overall, the findings suggest that web domain visibility is associated with domain-level research impact, as measured by the h-index, but does not directly translate into higher citation counts at the dataset level. While visibility may enhance discoverability, other factors play a crucial role in shaping citation impact. Furthermore, it is important to acknowledge that correlation does not imply causation. While the associations observed in this study suggest a relationship between visibility and impact, further investigation is required to understand the causal mechanisms underlying these patterns.

Many web domains lacked citations for their datasets, likely due to the widespread practice of informal data citation for data sharing and reuse in research papers, as highlighted by Park, You & Wolfram (2018). Their study in the biological/biomedical sciences field highlighted the prominence of informal data citations within the main text of articles, contrasting with the less frequent occurrence of formal data citations within references. Considering the significant number of datasets and web domains lacking citations, we performed correlation calculations under different conditions: first, without any filtering; second, by exclusively including web domains with an h-index greater than 0; and third, by also encompassing web domains with an h-index greater than one.

As a whole, the h-index of web domains for data repositories, as defined in this study, serves as a plausible indicator of the expected citation rate for datasets in the Social Sciences and Economics, including those not traditionally classified under these fields. The highly cited datasets tend to be clustered together. Most influential websites are also well-known in the community and offer plausible and useful contributions. Additionally, we have an indication that these findings likely extend to other forms of scholarly artifacts that are not always explicitly referred to as data. However, this aspect requires further investigation to be fully understood.

**Table 4. Correlation between datasets' number of received citations in *Social Sciences* and visibility index of web domain captured by Sistrix.**

| *Threshold* | *Number of Datasets* | *Spearman Correlation (P-Value)* |
|---|---|---|
| All datasets | 146,730 | 0.14 (0.0) |
| Datasets with number of citations > 0 | 7,183 | 0.31 (0.0) |
| Datasets with number of citations > 1 | 1,854 | 0.01 (0.633) |

**Table 5. Correlation between datasets' number of received citations in Economics and visibility index of web domain captured by Sistrix.**

| *Threshold* | *Number of Datasets* | *Spearman Correlation (P-Value)* |
|---|---|---|

---

[8] The code available here: https://github.com/momenifi/Dataset_finability/blob/main/3correlation_analysis.py

| | | |
|---|---|---|
| All datasets | 34,969 | 0.15 (0.0) |
| Datasets with number of citations > 0 | 2,932 | 0.37 (0.0) |
| Datasets with number of citations > 1 | 946 | -0.02 (0.555) |

**Conclusion and Discussion**

The fact that 99% of Social Sciences datasets and 98% of Economics datasets have no citations is a big challenge. Onyancha (2016) underscores the challenges in data citation compared to research articles, particularly in sub-Saharan Africa (SSA), where data citation rates are notably lower. It makes us wonder why these datasets aren't getting cited much. To address these challenges, increasing awareness among researchers about proper data citation is crucial. Establishing clear citation standards and providing incentives for proper attribution could encourage better citation practices. Additionally, improving metadata quality and repository infrastructure will enhance the discoverability of datasets, making them more accessible for researchers. Some researchers might be sharing data informally. Researchers may reuse data informally, without following the usual ways of citing it, such as the Joint Declaration of Data Citation Principles (Martone, 2014).

Also, some repositories do not make datasets easily findable, which impacts their citation rates. To fix this, they need to improve metadata standards to enhance dataset discoverability. We could make clear rules for citing data and give rewards to encourage people to do it properly, raising awareness among researchers in Social Sciences and Economics about proper data citation. Also, encourage repositories to optimize visibility through structured metadata and persistent identifiers (PIDs). we should make it easier to find these datasets by improving the information about them in the places they're stored.

Our correlation analysis suggests that datasets hosted on more visible web domains tend to receive more citations. At the dataset level, we find a positive correlation between dataset citation counts and web domain visibility, particularly for datasets with at least one citation. However, when analyzing domain-level citation metrics, such as the h-index, mean citation, and median citation, the correlations are less consistent. While higher visibility domains tend to host datasets with greater overall citation impact, the distribution of citations across datasets varies widely. Importantly, these correlations do not imply a causal relationship—higher domain visibility does not necessarily predict higher dataset citation rates. Instead, other factors, including dataset quality, repository policies, and researcher behaviors, play a role in shaping both visibility and impact.

To address these challenges and improve dataset findability and citation rates, repositories should consider implementing structured metadata, persistent identifiers, and FAIR Signposting. FAIR Signposting, a lightweight mechanism using standardized HTTP link headers, can enhance dataset discovery and usability by guiding researchers and automated tools toward relevant metadata, licensing information, and dataset relationships (Wilkinson et al., 2022). Integrating FAIR Signposting into repositories could:

- Make datasets easier to find by enabling automated tools and search engines to retrieve structured metadata.
- Improve citation tracking by linking datasets directly to related publications and persistent identifiers (PIDs).
- Facilitate better metadata interoperability, ensuring datasets are consistently indexed and referenced across different platforms.

Overall, while our findings suggest a relationship between repository visibility and citation impact, significant efforts are needed to improve dataset discoverability, formalize citation practices, and encourage researchers to attribute data properly. Addressing these challenges will be crucial in ensuring that datasets in Social Sciences and Economics gain the recognition and reuse they deserve.

**Limitation**

We acknowledge several limitations in this study. First, the challenge of studying data citation is evident, as many datasets in OpenAlex lack citations. It remains unclear whether these datasets receive fewer citations than published papers or if authors often rely on informal citations for data sharing and reuse (Park, You & Wolfram, 2018). This results in a skewed distribution of citation data, which may limit the accuracy of correlation analysis and lead to an underestimation of the impact of repository visibility on citation rates. The inaccessibility of informal citations may lead to incomplete correlations, limited insights, and potential underestimation of the impact on repository discoverability.

Additionally, while our analysis encompassed a wide array of datasets from OpenAlex, it's essential to note that our examination did not specifically quantify or track the prevalence of atypical datasets, such as online presentations in video format, within the dataset.

While these atypical datasets were part of our dataset, their specific prevalence or impact wasn't quantified or tracked within our analysis. Future research might benefit from quantifying these unconventional datasets' prevalence and impact on data citation practices.

**Data availability**

Data and code are accessible via the following link: https://github.com/momenifi/methodHub/blob/main/dataset_findability/

**Acknowledgements**

We gratefully acknowledge the German Competence Center for Bibliometrics (Grant No. 01PQ17001) for maintaining the dataset used in our analyses, as well as the financial support provided by BERD@NFDI (www.berd-nfdi.de, Grant No. 460037581).

**Appendix**

*Appendix A*

This appendix presents the top web domains based on various criteria. Tables are provided to showcase the top ten web domains with the highest number of DOIs, mean visibility index, and $h$-index in both the Economics and Social Sciences datasets.

**Table A.1: Top ten web domains with highest DOI numbers (Economics)**

| Domain | DOIs Number | Mean Visibility Index | h-Index |
|---|---|---|---|
| https://referenceworks.brillonline.com/ | 9,128 | 0.122337 | 74 |
| https://primarysources.brillonline.com/ | 8,941 | 0.000383 | 11 |
| https://www.socialscienceregistry.org/ | 5,676 | 0.000933 | 9 |
| https://psycnet.apa.org:443/ | 1,649 | 0.182410 | 8 |
| https://connect.h1.co/ | 1,639 | 0.000000 | 2 |
| https://www.oecd-ilibrary.org/ | 1,594 | 0.285173 | 12 |
| https://www.healthaffairs.org/ | 1,301 | 0.041977 | 3 |
| https://www.openicpsr.org/ | 1,212 | 0.001127 | 1 |
| https://www.iucnredlist.org/ | 858 | 0.073103 | 2 |
| https://www.degruyter.com/ | 576 | 2.044188 | 5 |

**Table A.2: Top ten web domains with highest DOIs (Social Sciences)**

| Domain | DOIs Number | Mean Visibility Index | h-Index |
|---|---|---|---|
| https://primarysources.brillonline.com/ | 45,967 | 0.000383 | 18 |
| https://referenceworks.brillonline.com/ | 26,343 | 0.122337 | 98 |
| https://scholarlyeditions.brill.com/ | 18,509 | 0.000000 | 1 |
| https://psycnet.apa.org:443/ | 16,820 | 0.182410 | 23 |
| https://connect.h1.co/ | 7,599 | 0.000000 | 9 |
| https://www.socialscienceregistry.org/ | 7,113 | 0.000933 | 6 |
| https://connect.liblynx.com/ | 4,715 | 0.000000 | 0 |
| https://www.degruyter.com/ | 2,649 | 2.044188 | 7 |
| https://www.healthaffairs.org/ | 1,961 | 0.041977 | 3 |
| https://doi.pangaea.de/ | 1,800 | 0.001448 | 4 |

**Table A.3: Top ten web domains with highest visibility index (Economics)**

| Domain | DOIs Number | Mean Visibility Index | h-Index |
|---|---|---|---|
| https://www.youtube.com/ | 2 | 61.709560 | 0 |
| https://www.researchgate.net/ | 8 | 4.404945 | 0 |
| https://onlinelibrary.wiley.com/ | 23 | 2.051243 | 1 |
| https://www.degruyter.com/ | 576 | 2.044188 | 5 |
| https://www.fao.org/ | 216 | 1.310798 | 1 |
| https://www.cambridge.org/ | 3 | 0.950633 | 0 |
| https://www.fs.usda.gov/ | 20 | 0.784208 | 1 |
| https://www.science.org/ | 166 | 0.770167 | 4 |
| https://www.ebi.ac.uk/ | 2 | 0.570627 | 0 |
| https://www.erudit.org/ | 1 | 0.448252 | 0 |

**Table A.4: Top ten web domains with highest visibility index (Social Sciences)**

| Domain | DOIs Number | Mean Visibility Index | h-Index |
|---|---|---|---|
| https://www.youtube.com/ | 4 | 61.709560 | 0 |
| https://link.springer.com/ | 5 | 6.462607 | 0 |
| https://www.researchgate.net/ | 28 | 4.404945 | 0 |
| https://www.jstor.org/ | 1 | 3.635153 | 0 |
| https://onlinelibrary.wiley.com/ | 86 | 2.051243 | 1 |
| https://www.degruyter.com/ | 2,649 | 2.044188 | 7 |
| https://www.fao.org/ | 271 | 1.310798 | 1 |
| https://www.cambridge.org/ | 2 | 0.950633 | 0 |
| https://www.fs.usda.gov/ | 370 | 0.784208 | 6 |
| https://www.science.org/ | 1,241 | 0.770167 | 10 |

**Table A.5: Top ten web domains with highest h-index (Economics)**

| Domain | DOIs Number | Mean Visibility Index | h-Index |
|---|---|---|---|
| https://referenceworks.brillonline.com/ | 9,128 | 0.122337 | 74 |
| https://www.oecd-ilibrary.org/ | 1,594 | 0.285173 | 12 |
| https://primarysources.brillonline.com/ | 8,941 | 0.000383 | 11 |
| https://www.socialscienceregistry.org/ | 5,676 | 0.000933 | 9 |
| https://psycnet.apa.org:443/ | 1,649 | 0.182410 | 8 |
| https://www.degruyter.com/ | 576 | 2.044188 | 5 |
| https://www.science.org/ | 166 | 0.770167 | 4 |
| https://www.healthaffairs.org/ | 1,301 | 0.041977 | 3 |
| https://www.authorea.com/ | 372 | 0.007120 | 3 |
| https://doi.pangaea.de/ | 107 | 0.001448 | 3 |

**Table A.6: Top ten web domains with highest h-index (Social Sciences)**

| Domain | DOIs Number | Mean Visibility Index | h-Index |
|---|---|---|---|
| https://referenceworks.brillonline.com/ | 26,343 | 0.122337 | 98 |
| https://psycnet.apa.org:443/ | 16,820 | 0.182410 | 23 |
| https://primarysources.brillonline.com/ | 45,967 | 0.000383 | 18 |
| https://www.oecd-ilibrary.org/ | 1,451 | 0.285173 | 11 |
| https://www.science.org/ | 1,241 | 0.770167 | 10 |
| https://connect.h1.co/ | 7,599 | 0.000000 | 9 |
| https://www.degruyter.com/ | 2,649 | 2.044188 | 7 |
| https://www.socialscienceregistry.org/ | 7,113 | 0.000933 | 6 |
| https://oxfordbibliographies.com/ | 418 | 0.000000 | 6 |
| https://www.fs.usda.gov/ | 370 | 0.784208 | 6 |